\RequirePackage{fix-cm}
\documentclass[smallcondensed]{svjour3}     
\smartqed  
\usepackage{graphicx}
\usepackage{mathtools}
\usepackage{tikz}
\usepackage{subfigure}
\usepackage{adjustbox}
\usepackage{setspace}

\begin{document}

\title{A Control Forced Concurrent Precursor Method for LES Inflow
}


\author{John S. Haywood         \and
        Adrian Sescu 
}


\institute{John S. Haywood \at
               Mississippi State University - Department of Aerospace Engineering\\330 Walker at Hardy Rd, Mississippi State, MS 39762\\
              \email{jsh478@msstate.edu}           
}

\date{Received: date / Accepted: date}

\maketitle
\begin{abstract}
With the increased application of large eddy simulation techniques, the generation of realistic turbulence at inflow boundaries is crucial for the accuracy of a simulation. The Control Forced Concurrent Precursor Method (CFCPM) proposed in this work combines an existing concurrent precursor method and a mean flow forcing method with a new extension of the controlled forcing method to impose turbulent inflow boundary conditions primarily, although not exclusively, for domains that require periodic boundary conditions. Turbulent inflow boundary conditions are imposed through a region of body forces added to the momentum equations of the main simulation that transfers the precursor simulation into the main domain. Controlled forcing planes, which come into play as body forces added to the momentum equations on planes perpendicular to the flow, located in the precursor simulation, allow for specific Reynolds stress tensors and mean velocities to be imposed. The mean flow controlled forcing method only modifies the mean velocity profiles, leaving the fluctuating velocity field untouched. The proposed fluctuating flow controlled forcing methods extends the application of the original controlled forcing method to multiple fluctuating velocity components and couples their calculation in order to amplify the existing fluctuations present in the precursor flow field so that prescribed anisotropic Reynolds stress tensors can be reproduced. The new method was tested on high Reynolds number turbulent boundary layer flow over a wall-mounted cube and low Reynolds number turbulent boundary layer flow over a backward-facing step. It was found that the new extension of the controlled forcing method reduced the development time for both test cases considered here when compared to not using controlled forcing and only using the original controlled forcing method.
\end{abstract}

\section{Introduction}\label{intro}
Precursor simulation methods have the ability of imposing the most realistic turbulent inflow conditions. The fluctuations imposed by a precursor simulation method are actual solutions to the Navier-Stokes equations and capture most of the higher order spatial and temporal statistics along with the physical dynamics associated with the coherent flow structures. This is in contrast to synthetic turbulence methods which forgo the explicit solution of the Navier-Stokes equations and impose artificial fluctuations that model certain statistics and/or coherent structures of realistic turbulence. The synthetic fluctuations are transformed into realistic turbulence by the Navier-Stokes equations over a certain development length which is usually one of the indicators of the model effectiveness. A concurrent precursor method (CPM) circumvents the large storage and file operation requirements of the database precursor methods by running the precursor simulation simultaneously with the main simulation. The recycling and rescaling family of precursor methods introduced by Spalart \cite{Spalart} and simplified by Lund et at. \cite{Lund} are restricted for use only in wall-bounded flows. The method samples a plane of data normal to the streamwise direction and rescales the inner and outer velocity profiles using appropriate similarity laws for each wall-normal region. Xiao et al. \cite{Xiao} later generalized the overall recycling/rescaling methodology to be applicable to non-equilibrium inflow conditions without a homogeneous direction by removing the similarity property-based rescaling and replacing it with a process that rescales the velocity fluctuations based on prescribed normal Reynolds stress profiles. The entire precursor domain is rescaled every certain amount of iterations. These rescaling procedures all occur outside of the solution of the governing equations.

Stevens et al. \cite{Stevens} proposed a concurrent precursor method for periodic domains that allowed for a precursor simulation to provide ``inflow'' conditions by blending a region of the precursor flow into the main domain.  The blending operation consists of copying the region of precursor flow into the main domain and then using a blending function to remove the discontinuity between the main flow and the copied flow. This operation occurs outside of the solution of the Navier-Stokes equations and only produces a $C^0$ continuous flow field which was found to introduce spurious oscillations in spectral and pseudo-spectral methods. Munters et al. \cite{Munters} later replaced the blending operation with a penalization region and expanded the method to account for time-dependent inflow freestream direction. The penalization region forces the main flow towards the  corresponding precursor flow in that region through the addition of body forces to the governing equations. Because the transfer of the precursor flow to the main domain occurs inside the solution the Navier-Stokes equations, there is a more physical transition from the main flow to the precursor flow in the main domain. A major drawback of this particular precursor method is that a long development time is required for the desired fully turbulent flow to be reached.

The controlled forcing method was introduced by Spille-Kohoff and Kaltenbach \cite{Spille_Kohoff} to accelerate the development of wall-bounded turbulence by increasing the production of Reynolds shear stress. This increase in production was achieved by adding body forces to the wall-normal momentum equation on planes normal to the flow that amplified the existing wall-normal velocity fluctuations. The amplitudes of the body forces were defined using a proportional-integral (PI) controller based on the error between given and calculated Reynolds shear stress profiles. Keating et al. \cite{Keating_1} showed that the addition of a controlled forcing method significantly reduced the development length for channel flow with turbulent inflow provided by a synthetic turbulence model. The wall-normal Reynolds stress was found by Laraufie et al. \cite{Laraufie} to provide more efficient control for the amplitude of the body forces than the Reynolds shear stress. 

In section \ref{numerical_method}, we briefly introduce the governing equations, the corresponding boundary conditions, and the numerical method utilized to discretize the equations. The Control Forced Concurrent Precursor Method (CFCPM) introduced in section \ref{section_CFCPM} combines the stationary concurrent precursor method of Munters et al. \cite{Munters} and the mean flow forcing of Schl\"uter et al. \cite{Schluter} with a new extension of the controlled forcing method. The proposed fluctuating flow controlled forcing extends the application of controlled forces to multiple fluctuating velocity components and couples their calculation to reproduce prescribed anisotropic Reynolds stress tensors. By applying the controlled forces to multiple fluctuating velocities and targeting multiple Reynolds stresses, the goal is to reduce the development time of the desired turbulent flow. Results from high and low Reynolds number turbulent boundary layer simulations are presented in section \ref{results} and are used to compare the ability of the proposed controlled forcing extension to decrease the development time of reproducing the desired turbulent conditions in the precursor flow with precursor flows without forcing and with only the original controlled forcing. The proposed forcing method is also compared at three different grid resolutions. The quality of the turbulent inflow conditions provided by the CFCPM is also investigated through comparisions with experimental data of the flow of these boundary layers over a wall-mounted cube and a backward-facing step, respectively.

\section{Governing Equations and Numerical Method}\label{numerical_method}
The governing equations are the filtered continuity and momentum equations for incompressible flow
\begin{equation}
\frac{\partial \tilde{u}_{i}}{\partial x_{i}} = 0
\end{equation}

\begin{equation}\label{CFCPM_momentum_eq}
\frac{\partial \tilde{u}_{i}}{\partial t}
+ \tilde{u}_{j}\frac{\partial \tilde{u}_{i}}{\partial x_{j}}
=
- \frac{\partial \tilde{p}^{*}}{\partial x_{i}}
- \frac{\partial \tilde{\tau}_{ij}}{\partial x_{j}}
+ \frac{1}{Re}\frac{\partial^2 \tilde{u}_{i}}{\partial x^2_j} 
+ 
 \begin{cases}
    (F_M)_i + (F_F)_i &;\ \text{precursor} \\
    (F_{CPM})_i + (F_{IBM})_i &;\ \text{main}
  \end{cases} 
\end{equation}
where spatial filtering is represented by a tilde, $\tilde{u}_{i}$ are the components of the velocity field corresponding to the streamwise $x_1$-direction, spanwise $x_2$-direction, and vertical $x_3$-direction, respectively, and $\tilde{p}^{*}$ is the effective pressure divided by the reference density. The SGS stress, $\tau_{ij}$, is modeled using the Lagrangian scale-dependent dynamic model developed by Bou-Zeid et al. \cite{Bou_Zeid}. Monin-Obukhov similarity theory is used to model the flow at the wall \cite{Bou_Zeid,Moeng,Monin}. The direct forcing immersed boundary method (IBM) is used to model the presence of a bluff body where the same Monin-Obukhov similarity theory is used at the surface of the body \cite{Mohd_Yusof,Tseng}. The immersed boundary method force, $(F_{IBM})_i$, is necessary because of the grid restrictions imposed by the pseudo-spectral method. There are three addition forcing terms associated with the CFCPM: the mean and fluctuating flow controlled forcing terms, $(F_M)_i$ and $(F_F)_i$, which maintain prescribed mean flow and turbulence levels in the precursor domain and the concurrent precursor method force, $(F_{CPM})_i$, which forces the main flow to match the precursor flow in the precursor forcing region. By adding the forcing terms to the momentum equations, before the solution of the Poisson equation for pressure, the forced flow is automatically divergence-free.

The numerical tool is a pseudo-spectral LES code that solves the filtered Navier-Stokes equations using a pseudo-spectral horizontal discretization and a centered finite difference staggered vertical uniform discretization \cite{Calaf,Sescu}. Time marching is performed using a fully-explicit second-order accurate Adams-Bashforth scheme \cite{Butcher}. The continuity equation is enforced through the solution of the Poisson equation resulting from taking the divergence of the momentum equation. Periodic boundary conditions are imposed along the horizontal directions. The CFCPM provides inflow boundary conditions that are introduced at the end of the main domain, which preserves the periodicity in the streamwise direction. The vertical gradients of velocity and the vertical velocity component vanish at the top boundary. The horizontal velocities at the first point away from the wall ($z = \Delta z/2$) are set through the velocity gradients in the vertical direction calculated using the Monin-Obukhov similarity theory and the vertical velocity at the wall is set to zero. The top and bottom boundary conditions are specific to turbulent boundary layer cases investigated in section \ref{results}.

\section{Control Forced Concurrent Precursor Method}\label{section_CFCPM}
\begin{figure}
 \begin{center}
   \begin{adjustbox}{width=10cm}
    \begin{tikzpicture}
	 
 	 \draw[thick] (0,3) rectangle (8,5);
 	 \draw[thick,red] (6.5,3) rectangle (8,5);
	 \draw[thick,blue] (0.75,3) -- (0.75,5);
	 \draw[thick,blue] (1,3) -- (1,5);
	 \draw[thick,blue] (1.25,3) -- (1.25,5);
	 \draw[thick,blue] (1.5,3) -- (1.5,5);
	 \draw[thick,blue] (1.75,3) -- (1.75,5);
	 \draw (-0.5,4) node[anchor=east] {\Large{Precursor Domain)}};

     \draw[<-,thick] (1.75,4) -- (2.75,4);
     \draw (4.2,4) node[anchor=south] {\Large{Controlled}};
     \draw (4.2,4) node[anchor=north] {\Large{Forcing Planes}};

	 \fill[red] (6.5,0) rectangle (8,2);
	 \draw[thick] (0,0) rectangle (8,2);
	 \draw (-0.5,1) node[anchor=east] {\Large{Main Domain)}};

     \draw[<-,thick] (7.5,1.5) arc (0:15:200pt);
     
     \draw[->,thick] (5.5,1) -- (6.5,1);
     \draw (4.2,1) node[anchor=south] {\Large{Precursor}};
     \draw (4.2,1) node[anchor=north] {\Large{Forcing Region}};
     
     \draw[->,thick] (2.5,2.5) node[anchor=east] {\Large{Flow}} -- (4.5,2.5);

    \end{tikzpicture}
   \end{adjustbox}
 \end{center}
  \caption{Example Control Forced Concurrent Precursor Method domains.}
 \label{example_domain}
\end{figure}
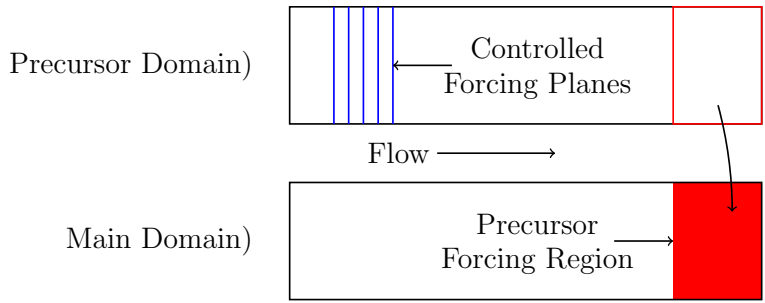

The Control Forced Concurrent Precursor Method is a combination of a traditional periodic concurrent precursor method with controlled forcing methods to allow for the imposition of given mean flow profiles and anisotropic Reynolds stress tensors. A region of the precursor flow field, outlined in red in figure \ref{example_domain}, is transfered to main domain through a forcing region, the red area in figure \ref{example_domain}, that penalizes the difference between the two flows. 

\subsection{Concurrent Precursor Method}\label{section_CPM}
The concurrent precursor method introduced by Stevens et al. \cite{Stevens} and modified by Munters et al. \cite{Munters} was specifically developed to provide inflow conditions for periodic domains, although it is not limited to periodic domains. A precursor domain is considered with identical dimensions and discretization as the main domain. The two simulations are carried out at the same flow conditions and are synchronized in time. Body forces, added to each of the momentum equations over a region of the main domain, force the main flow towards the precursor flow. They are defined as follows
\begin{eqnarray}
\left(F_{CPM}\right)_i(x,y,z,t) &=& \sigma(x) \left[ u_i^{pre}(x,y,z,t) - u_i^{main}(x,y,z,t) \right]\\
\sigma(x) &=& 
  \begin{cases}
    \sigma_{max} \left( \frac{x-x_s}{x_{pl}-x_s}\right)^n &;\ x_s \leq x \leq x_{pl} \\
    \sigma_{max}  & ;\ x_{pl} < x \leq L_x 
  \end{cases} 
\end{eqnarray}
where $\sigma$ is the precursor forcing strength, $x_s$ is the streamwise location of the start of the precursor forcing region and $x_{pl}$ is the streamwise location of the end of the increasing precursor forcing strength region. $x_{pl}$ and the exponent $n$ determine the smoothness of the transition and $\sigma_{max}$ is the maximum precursor forcing strength. The parameters used in this work are as follows:
\begin{eqnarray}
  x_s &=& 0.8 L_x \\ \nonumber
  x_{pl} &=& 0.99 L_x \\ 
  n &=& 5 \\ \nonumber
  \sigma_{max} &=& \frac{0.7}{\Delta t} \nonumber
\end{eqnarray}
where $L_x$ is the length of the flow domain and $\Delta t$ is the time step. The precursor forcing region can be located anywhere in the main domain where ``inflow'' conditions are desired.

The precursor and main domains are not required to have the same dimensions, grid resolutions, and/or flow conditions. These differences between the two domains introduce various complexities that must then be addressed (e.g., proper interpolation needs to be in place). Streamwise periodicity in the main domain can be introduced by using a precursor domain with a smaller streamwise length \cite{Nikitin}. In the case that the precursor and main domains do not share the same discretization, interpolation would be required to transfer the precursor flow to the main domain. For flow conditions that differ between the precursor and main domains, the precursor flow would also need to be rescaled to match the flow conditions in the main domain before it is transfered. Wu \cite{Wu} provides a review of the techniques developed in the context of the database precursor methods to correct for using dissimilar precursor domains and/or flows which could be applied in this method.

\subsection{Fluctuating Flow Controlled Forcing}\label{Fluctuation_Controlled_Forcing}
The new fluctuating flow controlled forcing extends the original method introduced by Spille-Kohoff and Kaltenbach \cite{Spille_Kohoff} to two and three dimensions and moves the calculation of the forces into the local principal-axis coordinate system in order to match 2D and 3D anisotropic Reynolds stress tensors. The fluctuating flow controlled forcing method only modifies fluctuations that already exist. It does not generate its own fluctuations. These existing velocity fluctuations are modified through body forces, $(F_F)_i$, added to the momentum equations on planes normal to the streamwise direction. The body forces are defined in the local principal-axis coordinate system as follows
\begin{equation}
\left(F_F\right)^p_i\left(x_f,y,z,t\right) = r^p_i\left(x_f,y,z,t\right) \left[u^p_i\left(x_f,y,z,t\right) - \langle u^p_i \rangle \left(x_f,y,z,t\right)\right]
\end{equation}
where $x_f$ are the streamwise locations of the fluctuating flow controlled forcing planes, $r^p_i$ are the amplitudes of the forces, $u^p_i$ are the instantaneous velocities at time $t$, and $\langle u^p_i \rangle$ are the mean velocities at time $t$. The superscript $^p$ denotes variables in the local principal-axis coordinate system. The amplitudes of the forces are defined using a PI controller based on the error between the given and calculated principal-axis Reynolds stress profiles.
\begin{eqnarray}
r^p_i\left(x_f,y,z,t\right) &=& \alpha_i \thinspace e^p_i\left(x_f,y,z,t\right) + \beta_i \int_{0}^{t} e^p_i\left(x_f,y,z,t'\right)dt' \\
e^p_i\left(x_f,y,z,t\right) &=&  \langle u_i'u_i' \rangle^p_{given} \left(x_f,y,z,t\right) - \langle u_i'u_i' \rangle^p \left(x_f,y,z,t\right)
\end{eqnarray}
$\alpha_i$ and $\beta_i$ are chosen such that the error decreases sufficiently fast without introducing numerical instabilities. In this work, $\alpha_i=1$ and $\beta_i=100$. $\langle u_i'u_i' \rangle^p_{given}$ and $\langle u_i'u_i' \rangle^p$ are the given and current principal-axis Reynolds stress profiles. Using the transformation matrix created from the eigenvectors of the local Reynolds stress tensor, $T^G_p$, the body forces are transformed from the local principal-axis coordinate system to the global coordinate system before being applied to the momentum equations.
\begin{eqnarray}
\left(F_F\right)_i = \left( T^G_p \right)_{ij} \left(F_F\right)^p_j
\end{eqnarray}

The following criteria for the application of the body forces ensure that the more energetic fluctuations are amplified consistent with the sign of cross-velocity correlations, while unrealistically large fluctuations are not amplified \cite{Keating_1,Spille_Kohoff}. 
\begin{eqnarray}
|u_i'| &<& 0.4\thinspace U_{\infty} \\ \nonumber
|u_1'u_3'| &>& 0.0015\thinspace U^2_{\infty}
\end{eqnarray}
If these criteria are not met,
\begin{equation}
\left(F_F\right)_i\left(x_f,y,z,t\right) = 0
\end{equation}
Keating et al. \cite{Keating_1} suggested using an exponential weighted moving average to calculate the current time-averaged quantities,
\begin{equation}
\langle \phi \rangle \left(t+\Delta t \right) = \phi\left(t\right)\frac{\Delta t}{T_{avg}} + \left( 1 - \frac{\Delta t}{T_{avg}} \right) \langle \phi \rangle \left(t\right)
\end{equation}
where $\phi$ is any quantity needing time-averaging, $\Delta t$ is the time step, and $T_{avg}$ is the averaging time period. In this work, $T_{avg}$ is set to two flow-throughs.

For a 2D Reynolds stress tensor, a force is not applied to the fluctuating velocity component where the associated Reynolds stresses are not known. If only the normal Reynolds stresses are known, the proposed method reduces to applying the original controlled forcing method to each velocity component independently. This method is also not restricted to a certain type of flow or numerical method. In principle, the controlled forcing method allows for time-varying Reynolds stresses. The force applied at a point is only based on the current and given Reynolds stresses. For flows that involve faster varying Reynolds stresses, care needs to be taken on how the current Reynolds stresses are determined such that a proper average can be calculated.

\subsection{Mean Flow Controlled Forcing}
The imposition of given mean flow profiles is handled in the same manner as Schl\"uter et al. \cite{Schluter}. Body forces, $(F_M)_i$, are added to the momentum equations in the precursor domain on planes normal to the streamwise direction in order to drive the mean flow towards a given mean profile. For a constant density flow, these body forces take the form of the following proportional controller
\begin{equation}
\left(F_M\right)_i\left(x_m,y,z,t\right) = \gamma_i \left[ \langle u_i \rangle_{given}\left(x_m,y,z,t\right) - \langle u_i \rangle \left(x_m,y,z,t\right) \right]
\end{equation}
where $x_m$ are the streamwise locations of the mean flow forcing planes, $\gamma_i$ are the mean flow forcing strength factors, $\langle u_i \rangle_{given}$ are the given mean velocity profiles, and $\langle u_i \rangle$ are the current mean velocity profiles at time $t$ calculated using the exponential weighted moving averaging shown in section \ref{Fluctuation_Controlled_Forcing}. The mean flow forcing strength factor should be large enough that the body forces promptly respond to changing flow conditions while also not being too large as to introduce numerical instabilities. In this work, $\gamma_i=0.7$. In the same manner as the fluctuating flow controlled forcing, the given mean velocities are allowed to be time-varying with the same caveat when calculating the current mean velocity for faster varying flows.

\section{Results and Discussion}\label{results}
Two validation cases are considered: a high and low Reynolds number turbulent boundary layer. The high Reynolds number case utilizes 3D Reynolds stress tensor profiles, while the low Reynolds number case only prescribes 2D tensor profiles. The objectives of these numerical simulations are to impose a experimentally measured turbulent boundary layer and then validate the LES results with the experimental results. For both cases, the proposed fluctuating flow controlled forcing method is compared with simulations without controlled forcing and with only the original controlled forcing of Spille-Kohoff and Kaltenbach \cite{Spille_Kohoff}. Additionally, the ability of the proposed method to match the given Reynolds stresses is evaluated by running the precursor simulation at three different grid resolutions.

\subsection{High Reynolds Number Turbulent Boundary Layer} \label{results_high}
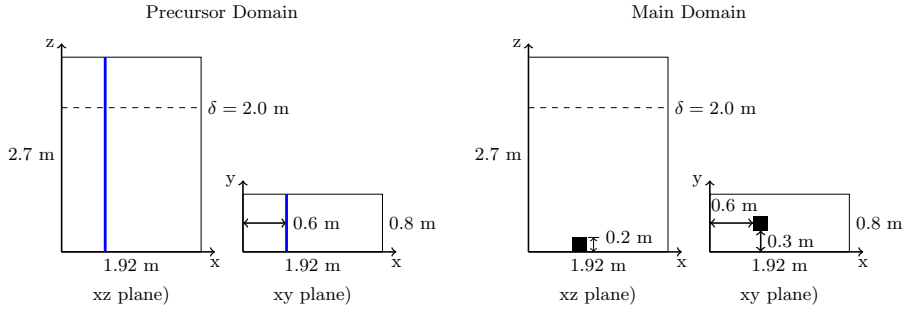
\begin{figure}[h]
 \begin{center}
 \begin{subfigure}{}
   \begin{adjustbox}{width=5.8cm}
    \begin{tikzpicture}

     \draw (2.875,4.1) node[anchor=south] {Precursor Domain};

     \draw (0,0) rectangle (2.5,3.515625);     
     \draw (1.25,0) node[anchor=north] {$1.92$ m};
     \draw (1.25,-0.5) node[anchor=north] {xz plane)};
     \draw (0,1.7578125) node[anchor=east] {$2.7$ m};
     \draw[->,thick] (0,0) -- (2.75,0) node[anchor=north] {x};
     \draw[->,thick] (0,0) -- (0,3.765625) node[anchor=east] {z};
     
     \draw[line width = 0.5mm,blue] (0.78125,0) -- (0.78125,3.515625);
     
     \draw[dashed] (0,2.60415) -- (2.5,2.60415);
     \draw (2.5,2.60415) node[anchor=west] {$\delta = 2.0$ m};
     
     \draw (3.25,0) rectangle (5.75,1.0415);
     \draw (4.5,0) node[anchor=north] {$1.92$ m};
     \draw (4.5,-0.5) node[anchor=north] {xy plane)};
     \draw (5.75,0.52075) node[anchor=west] {$0.8$ m};
     \draw[->,thick] (3.25,0) -- (6,0) node[anchor=north] {x};
     \draw[->,thick] (3.25,0) -- (3.25,1.2915) node[anchor=east] {y};
     
     \draw[<->,thick] (3.25,0.52075) -- (4.03125,0.52075);
     \draw (4.03125,0.52075) node[anchor=west] {$0.6$ m};
     \draw[line width = 0.5mm,blue] (4.03125,0) -- (4.03125,1.0415);
     
    \end{tikzpicture}
   \end{adjustbox}
 \end{subfigure}
 \begin{subfigure}{}
   \begin{adjustbox}{width=5.8cm}
    \begin{tikzpicture}

     \draw (2.875,4.1) node[anchor=south] {Main Domain};

     \draw (0,0) rectangle (2.5,3.515625);     
     \draw (1.25,0) node[anchor=north] {$1.92$ m};
     \draw (1.25,-0.5) node[anchor=north] {xz plane)};
     \draw (0,1.7578125) node[anchor=east] {$2.7$ m};
     \draw[->,thick] (0,0) -- (2.75,0) node[anchor=north] {x};
     \draw[->,thick] (0,0) -- (0,3.765625) node[anchor=east] {z};
     
     \fill[black] (0.78125,0) rectangle (1.0415,0.260415);
     \draw[<->] (1.1665,0) -- (1.1665,0.260415);
     \draw[-] (1.0665,0.260415) -- (1.2665,0.260415);
     \draw (1.2665,0.260415) node[anchor=west] {$0.2$ m};
     
     \draw[dashed] (0,2.60415) -- (2.5,2.60415);
     \draw (2.5,2.60415) node[anchor=west] {$\delta = 2.0$ m};
     
     \draw (3.25,0) rectangle (5.75,1.0415);
     \draw (4.5,0) node[anchor=north] {$1.92$ m};
     \draw (4.5,-0.5) node[anchor=north] {xy plane)};
     \draw (5.75,0.52075) node[anchor=west] {$0.8$ m};
     \draw[->,thick] (3.25,0) -- (6,0) node[anchor=north] {x};
     \draw[->,thick] (3.25,0) -- (3.25,1.2915) node[anchor=east] {y};
     
     \fill[black] (4.03125,0.3905625) rectangle (4.2915,0.6509375);
     \draw[<->,thick] (3.25,0.52075) -- (4.03125,0.52075);
     \draw (3.690625,0.62075) node[anchor=south] {$0.6$ m};
     \draw[<->,thick] (4.161375,0) -- (4.161375,0.3905625);
     \draw (4.161375,0.19528125) node[anchor=west] {$0.3$ m};
     
    \end{tikzpicture}
   \end{adjustbox}
 \end{subfigure}

 \end{center}
  \caption{The physical dimensions of the domain used for comparison to the experimental results of Castro and Robins\cite{Castro}. The blue line represents the single controlled forcing plane.}
 \label{Castro_domain}
\end{figure}

A high Reynolds number boundary layer flow developing over a $0.2$ m wall-mounted cube is considered at $Re_\theta=3.0\times10^5$. This was the subject of the wind tunnel experiment of Castro and Robins \cite{Castro}. The streamwise, spanwise, and vertical dimensions of domain are $1.92 \times 0.8 \times 2.7$ m, where the height of both the domain and the cube are equal to the height of the wind tunnel test section and the experimental cube. The precursor and main domains are shown in figure \ref{Castro_domain}, where the blue line represents the controlled forcing plane. The precursor and main domain are both periodic and use the same grid resolution. The three uniform discretizations considered are shown in table \ref{Castro_Grid_Cases}. The three fluctuating forcing methods (none, original, and proposed) are compared using the fine grid. 

A turbulent boundary layer of thickness $2.0$ m with a freestream velocity of $2.02$ m/s is imposed in the precursor simulation by a single controlled forcing plane using mean streamwise velocity and complete Reynolds stress tensor profiles given by Castro and Robins \cite{Castro}. For the original forcing method, only $\langle w'w' \rangle$ is used as a target. The initial fluctuations are provided by a random field of white noise scaled to match the Reynolds stress tensor profiles. The simulations comparing the forcing methods were all run until the first method converged to the desired Reynolds stresses, within $250$ flow-throughs; then statistics were collected for $30$ flow-throughs.

\begin{table}[htpb]
 \begin{center}
  \begin{tabular}{| c || c |} \hline
       Case	& $N_x \times N_y \times N_z$		\\	\hline \hline
       Coarse	& $64 \times 32 \times 64$\\	\hline
       Medium	& $128 \times 48 \times 128$\\	\hline
       Fine	& $192 \times 64 \times 324$\\
\hline
  \end{tabular}
    \caption{Grid point dimensions of the domain.}
  \label{Castro_Grid_Cases}
 \end{center}
\end{table}

\begin{figure}[h]
 \begin{center}
  \includegraphics[width=11cm, clip=true]{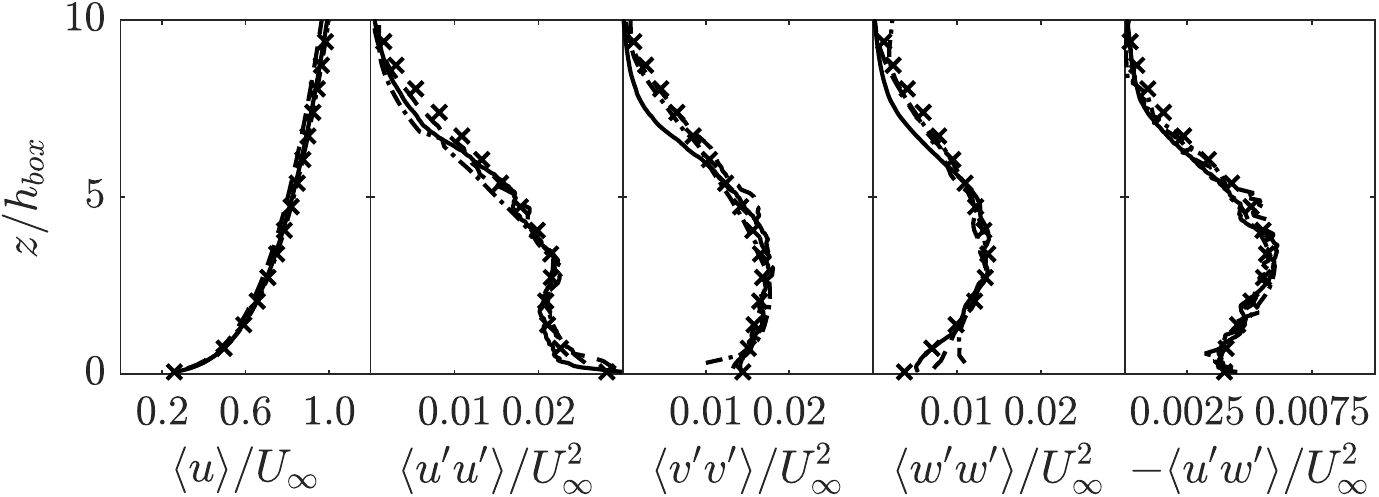}
 \end{center}
\caption{Turbulent statistics along the vertical direction from the precursor domain, using the proposed forcing method and three grid resolutions, sampled at the controlled forcing plane at conditions matching the momentum thickness Reynolds number of Castro and Robbins \cite{Castro}. -.-.-) coarse; -\ -\ -) medium; -----) fine; X) experimental data}
\label{castro_pre_grid}
\end{figure}

The mean streamwise velocity and non-zero Reynolds stress profiles sampled at the location of controlled forcing plane are plotted in figure \ref{castro_pre_grid}. These profiles were collected from the precursor simulation using the proposed forcing method for three different grid resolutions and compared with the experimental data from Castro and Robins \cite{Castro}. The proposed forcing method was able to successfully match the target Reynolds stress profiles for all three grid resolutions. This is not surprising because the controlled forcing method operates on each discrete grid point on the controlled forcing plane independently. The forces associated with a point on the forcing plane are only controlled by the local velocity fluctuations and target values at that point.

\begin{figure}[h]
 \begin{center}
 Precursor) \\
  \includegraphics[width=12cm, clip=true]{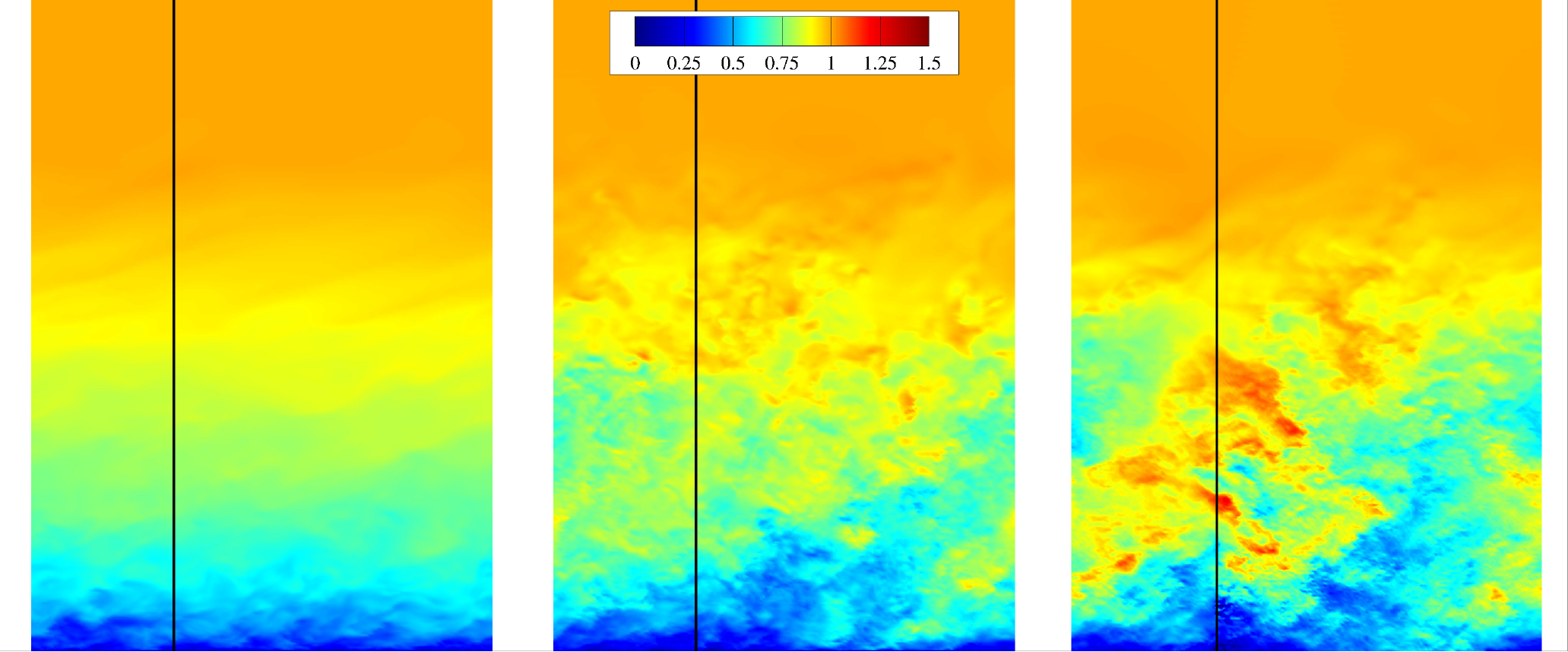} \\ \vspace{4mm} 
 Main) \\
  \includegraphics[width=12cm, clip=true]{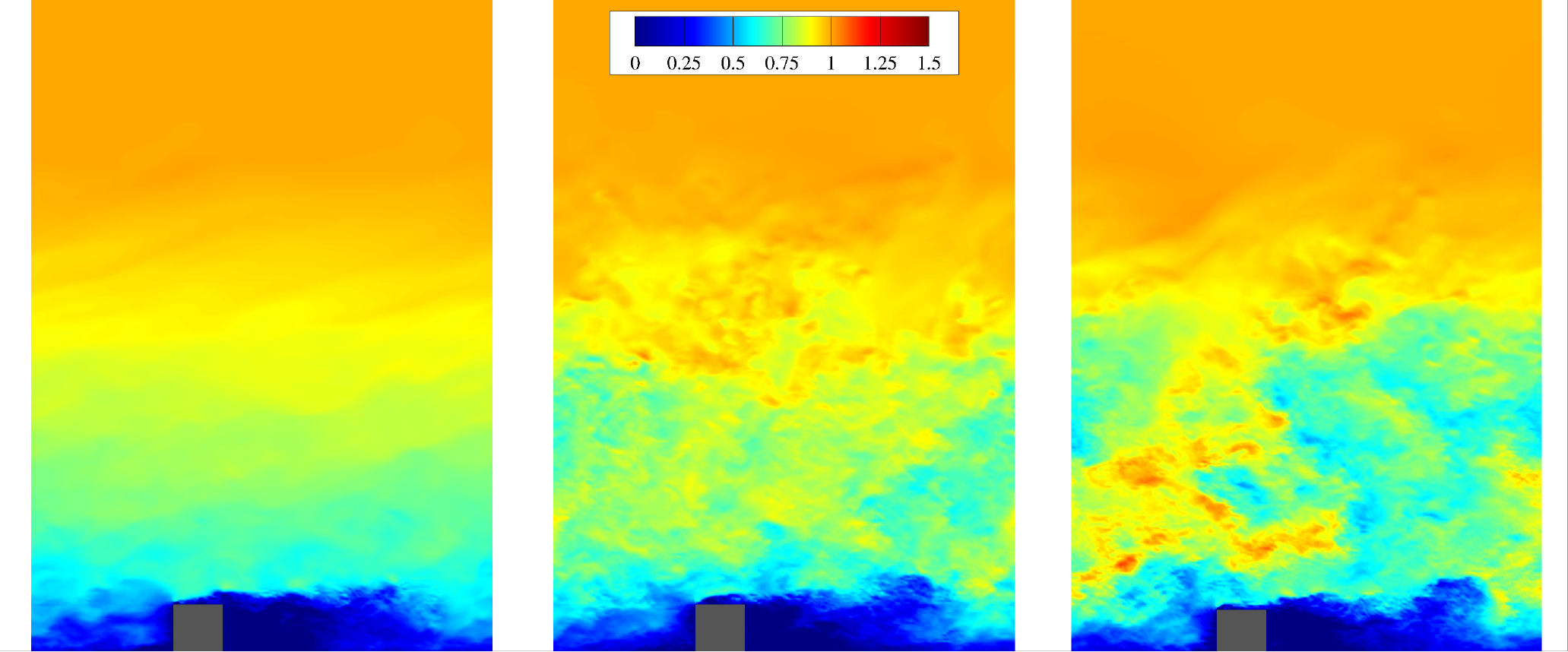}
 \end{center}
\caption{Contour of the instantaneous streamwise velocity normalized by the freestream velocity on an xz plane through the center of the cube, at conditions matching the momentum thickness Reynolds number in Castro and Robins \cite{Castro}. The black line represents the controlled forcing plane.} top) precursor domain; bottom) main domain; left) ``no forcing''; middle) ``original forcing''; right) proposed forcing
\label{castro_u_contour}
\end{figure}

Contour plots of the instantaneous streamwise velocity in the precursor and main domain for the three different forcing cases are presented in figure \ref{castro_u_contour}. For all three forcing cases, the flow structures immediately upstream of the outlet and downstream of inlet are extremely similar when comparing their respective main and precursor domains. Also notice how the wake created by the cube in the main domain is smoothly transformed by the precursor forcing region into the precursor flow field. When comparing the different controlled forcing methods present in the precursor simulations, the level of turbulence present in the precursor simulations increases moving from the ``no forcing'' case on the left to the proposed forcing case on the right. Through the addition of controlled forces on more of the fluctuating velocity components, the development of the turbulent boundary layer is accelerated. It is worth noting that the controlled forcing plane (the black line on the precursor contours) only affects the intensity of turbulent structures as it passes through it, and does not change the shape of the structures. In the main domain contours, the box wakes are all similar in shape below the height of the box.

\begin{figure}[h]
 \begin{center}
  \includegraphics[width=11cm, clip=true]{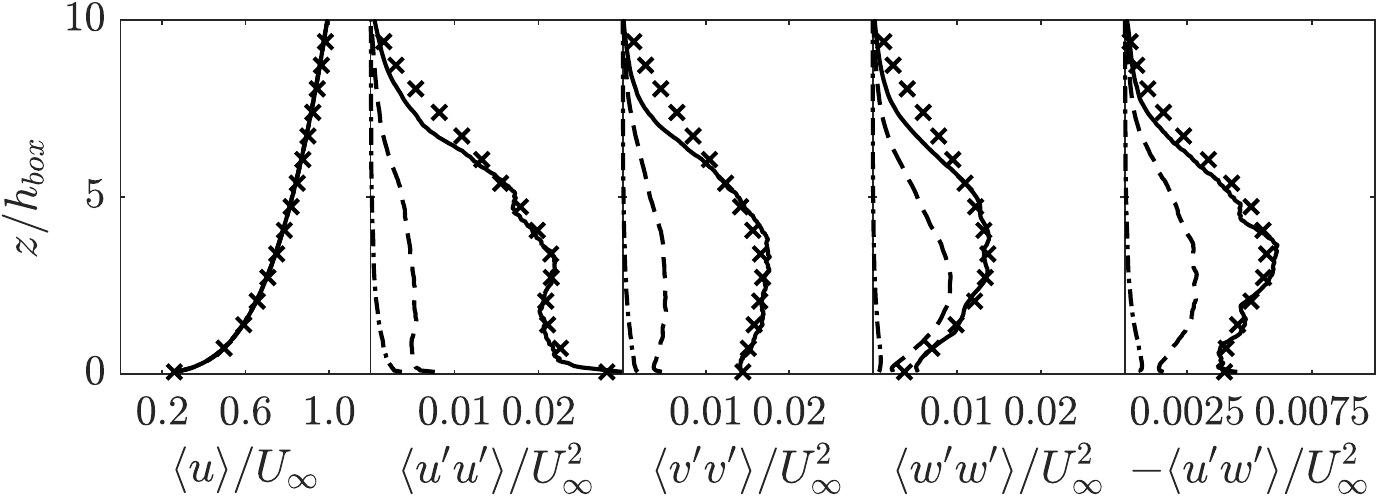}
 \end{center}
\caption{Turbulent statistics along the vertical direction from the precursor domain sampled at the controlled forcing plane at conditions matching the momentum thickness Reynolds number of Castro and Robbins \cite{Castro}. -.-.-) ``no forcing''; -\ -\ -) ``original forcing''; -----) proposed forcing; X) experimental data }
\label{castro_pre}
\end{figure}

The mean streamwise velocity and non-zero Reynolds stress profiles from the precursor simulation are plotted in figure \ref{castro_pre}, alongside the experimental data from Castro and Robins\cite{Castro} at the location of controlled forcing plane. In the computational time it took for the proposed forcing method to grow the desired turbulent boundary layer and match all of the Reynolds stress profiles, the boundary layers in the ``no forcing'' and ``original forcing'' cases were still developing and hence had not matched the target Reynolds stresses yet. Without forcing, the turbulence produced at the bottom wall is left to naturally grow into the domain. This is evidenced by slightly larger Reynolds stresses close to the wall, quickly decreasing to near zero moving towards the freestream, and the lack of turbulent structures seen away from the wall in the top-left contour of figure \ref{castro_u_contour}. Because the original forcing method actively targeted the wall-normal Reynolds stress, it does show an increased level. The Reynolds shear stress also shows an increase, which is consistent with the primary function of the original controlled forcing method. These increases agrees with what is seen in the top-middle contour of figure \ref{castro_u_contour}; turbulent structures are present, but they have a lower magnitude than the structures seen in the top-right contour of figure \ref{castro_u_contour}. If the ``no forcing'' and ``orginal forcing'' simulations were allowed to continue integrating the governing equations, it is expected that the desired turbulent boundary layer would eventually develop. The comparisons in figure \ref{castro_pre} shows that the proposed forcing method reduces the development time of the turbulent boundary layer as compared to the ``no forcing'' and ``original forcing'' cases, thus reducing the overall computational cost.

\begin{figure}[h]
 \begin{center}
  \includegraphics[width=8.25cm, clip=true]{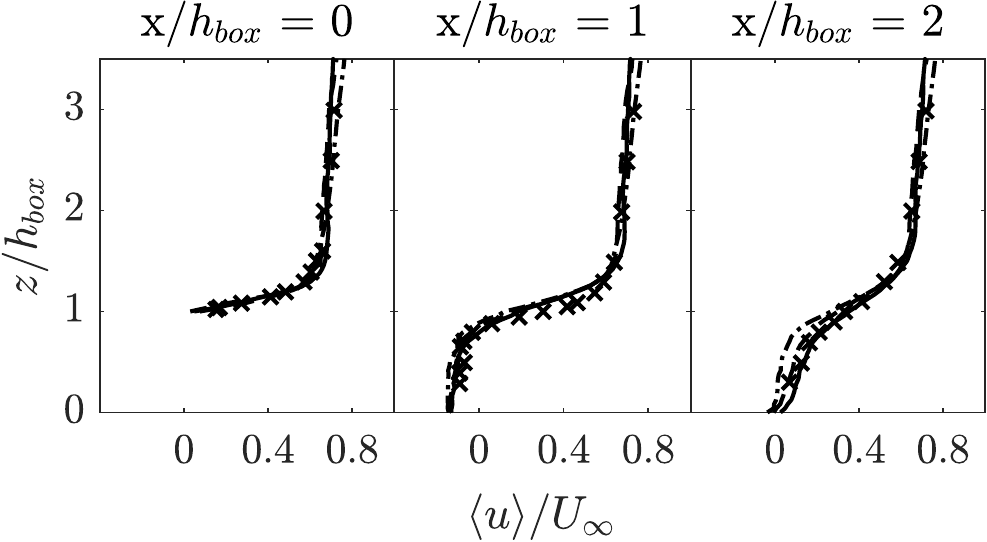} \\ \vspace{4mm}
  \includegraphics[width=8.25cm, clip=true]{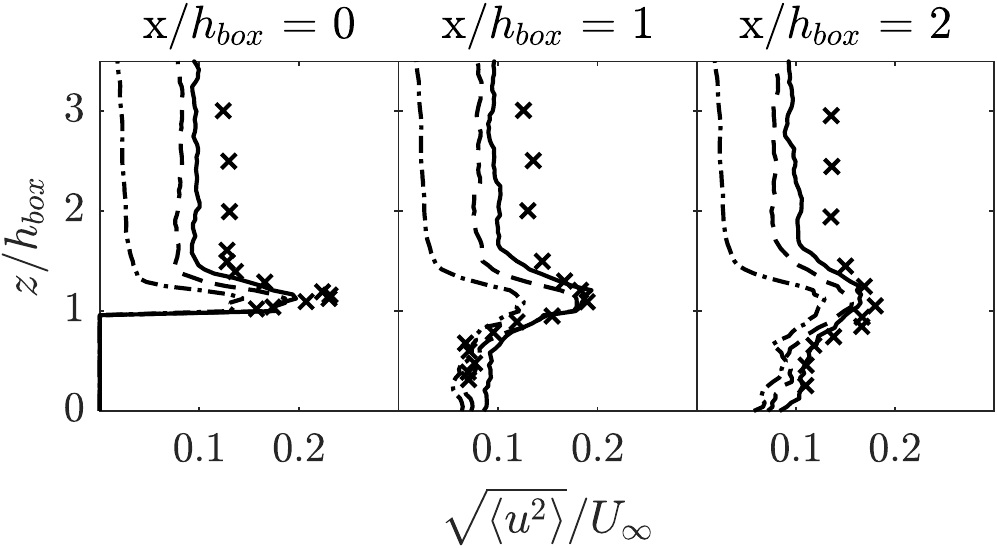}
 \end{center}
\caption{Turbulent statistics along the vertical direction sampled at streamwise locations downstream of the cube. -.-.-) ``no forcing''; -\ -\ -) ``original forcing''; -----) proposed forcing; X) experimental data for the high Reynolds number case of Castro and Robbins \cite{Castro}; top) Mean streamwise velocity; bottom) Streamwise turbulence intensity.    }
\label{castro_main}
\end{figure}

Vertical profiles of the mean streamwise velocity and turbulence intensity from the main simulation are plotted in figure \ref{castro_main} alongside the experimental data from Castro and Robins at three streamwise locations in the wake, $x/h_{box}= \{0,1,2\}$. Good agreement was found for the mean streamwise velocity for all three methods. In terms of the turbulence intensity, the proposed forcing method showed the best agreement. The other two forcing methods both underpredicted the turbulence intensity above the height of the box, which is consistent with the low intensity turbulence in their precursor simulations. The effect of the ``original forcing'' method is seen in the increased turbulence intensity as compared to the ``no forcing'' case. It is worth noting that below the height and at one box height downstream of the cube all three cases give reasonable intensities, which supports the similar wake shape seen in the bottom contours of figure \ref{castro_u_contour}.  This suggests that immediately behind the cube, the presence of the cube and the correct mean flow are more significant than the correct levels of freestream turbulence.

\subsection{Low Reynolds Number Turbulent Boundary Layer} \label{results_low}
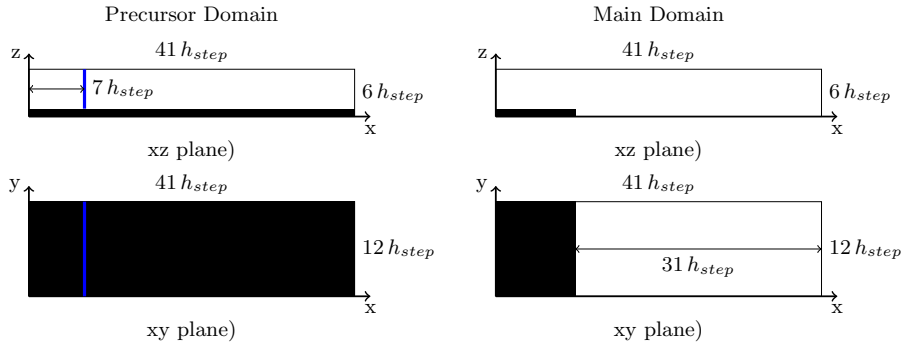
\begin{figure}[h]
 \begin{center}
 \begin{subfigure}{}
   \begin{adjustbox}{width=5.8cm}
    \begin{tikzpicture}

     \draw (2.5625,4.1) node[anchor=south] {Precursor Domain};

     \draw (0,2.7) rectangle (5.125,3.45);     
     \draw (2.5625,3.45) node[anchor=south] {$41\thinspace h_{step}$};
     \draw (2.5625,2.45) node[anchor=north] {xz plane)};
     \draw (5.125,3.075) node[anchor=west] {$6\thinspace h_{step}$};
     \draw[->,thick] (0,2.7) -- (5.375,2.7) node[anchor=north] {x};
     \draw[->,thick] (0,2.7) -- (0,3.7) node[anchor=east] {z};
     
     \fill[black] (0,2.7) rectangle (5.125,2.825);
     
     \draw[line width = 0.5mm,blue] (0.875,2.825) -- (0.875,3.45);
     \draw[<->] (0,3.1375) -- (0.875,3.1375);
     \draw (0.875,3.1375) node[anchor=west] {$7\thinspace h_{step}$};
     
     \draw (0,-0.15) rectangle (5.125,1.35);     
     \draw (2.5625,1.35) node[anchor=south] {$41\thinspace h_{step}$};
     \draw (2.5625,-0.4) node[anchor=north] {xy plane)};
     \draw (5.125,0.6) node[anchor=west] {$12\thinspace h_{step}$};
     \draw[->,thick] (0,-0.15) -- (5.375,-0.15) node[anchor=north] {x};
     \draw[->,thick] (0,-0.15) -- (0,1.6) node[anchor=east] {y};
     
     \fill[black] (0,-0.15) rectangle (5.125,1.35);
     \draw[line width = 0.5mm,blue] (0.875,-0.15) -- (0.875,1.35);

    \end{tikzpicture}
   \end{adjustbox}
 \end{subfigure}
  \begin{subfigure}{}
   \begin{adjustbox}{width=5.8cm}
    \begin{tikzpicture}

     \draw (2.5625,4.1) node[anchor=south] {Main Domain};

     \draw (0,2.7) rectangle (5.125,3.45);     
     \draw (2.5625,3.45) node[anchor=south] {$41\thinspace h_{step}$};
     \draw (2.5625,2.45) node[anchor=north] {xz plane)};
     \draw (5.125,3.075) node[anchor=west] {$6\thinspace h_{step}$};
     \draw[->,thick] (0,2.7) -- (5.375,2.7) node[anchor=north] {x};
     \draw[->,thick] (0,2.7) -- (0,3.7) node[anchor=east] {z};
     
     \fill[black] (0,2.7) rectangle (1.25,2.825);
     
     \draw (0,-0.15) rectangle (5.125,1.35);     
     \draw (2.5625,1.35) node[anchor=south] {$41\thinspace h_{step}$};
     \draw (2.5625,-0.4) node[anchor=north] {xy plane)};
     \draw (5.125,0.6) node[anchor=west] {$12\thinspace h_{step}$};
     \draw[->,thick] (0,-0.15) -- (5.375,-0.15) node[anchor=north] {x};
     \draw[->,thick] (0,-0.15) -- (0,1.6) node[anchor=east] {y};
     
     \fill[black] (0,-0.15) rectangle (1.25,1.35);
     \draw[<->] (1.25,0.6) -- (5.125,0.6);
     \draw (3.1875,0.6) node[anchor=north] {$31\thinspace h_{step}$};
    \end{tikzpicture}
   \end{adjustbox}
 \end{subfigure}
 \end{center}
  \caption{The physical dimensions of the domain used for comparison to the experimental results of Jovic and Driver. The blue line represents the single controlled forcing plane.}
 \label{step_domain}
\end{figure}

A low Reynolds number turbulent boundary layer flow developing over a $0.0098$ m backward-facing step is considered at $Re_\theta = 6.1 \times 10^2$. Jovic and Driver \cite{Jovic} carried out this experiment in a wind tunnel as a companion to the DNS validation of Le et al. \cite{Le}. The streamwise, spanwise, and vertical dimensions of the domain are $41 h_{step} \times 12 h_{step} \times 6 h_{step}$, where the height of the step is equal to the height of the experimental step and the size of the domain is comparable to the size of the numerical domain in Le et al. \cite{Le}. The precursor and main domains are shown in figure \ref{step_domain}, where the blue line represents the controlled forcing plane. The precursor and main domain are both periodic and use the same grid resolution. The three uniform discretizations considered are shown in table \ref{Step_Grid_Cases}. The three forcing methods (none, original, and proposed) are compared using the fine grid.
\begin{table}[htpb]
 \begin{center}
  \begin{tabular}{| c || c |} \hline
       Case	& $N_x \times N_y \times N_z$	\\	\hline \hline
       Coarse	& $128 \times 32 \times 96$	\\	\hline
       Medium	& $256 \times 64 \times 144$	\\	\hline
       Fine	& $384 \times 96 \times 192$	\\
\hline
  \end{tabular}
    \caption{Grid point dimensions of the low Reynolds number turbulent boundary layer domain.}
  \label{Step_Grid_Cases}
 \end{center}
\end{table}
A turbulent boundary layer of thickness $0.0115$ m with a freestream velocity of $7.72$ m/s is imposed in the precursor simulation by a single controlled forcing plane using mean streamwise velocity and 2D Reynolds stress tensor profiles given by Jovic and Driver \cite{Jovic}. Because correlations are only known for the streamwise and vertical velocities, no force was applied to the spanwise momentum equation for the proposed forcing method. As in section \ref{results_high}, only $\langle w'w'\rangle$ is used as a target for the original forcing method. The initial fluctuations are provided by a random field of white noise scaled to match the Reynolds stress tensor profiles. The simulations comparing the forcing methods were all run until the first method converged to the desired Reynolds stresses, within $144$ flow-throughs; then statistics were collected for $30$ flow-throughs.

\begin{figure}[h]
 \begin{center}
  \includegraphics[width=11cm, clip=true]{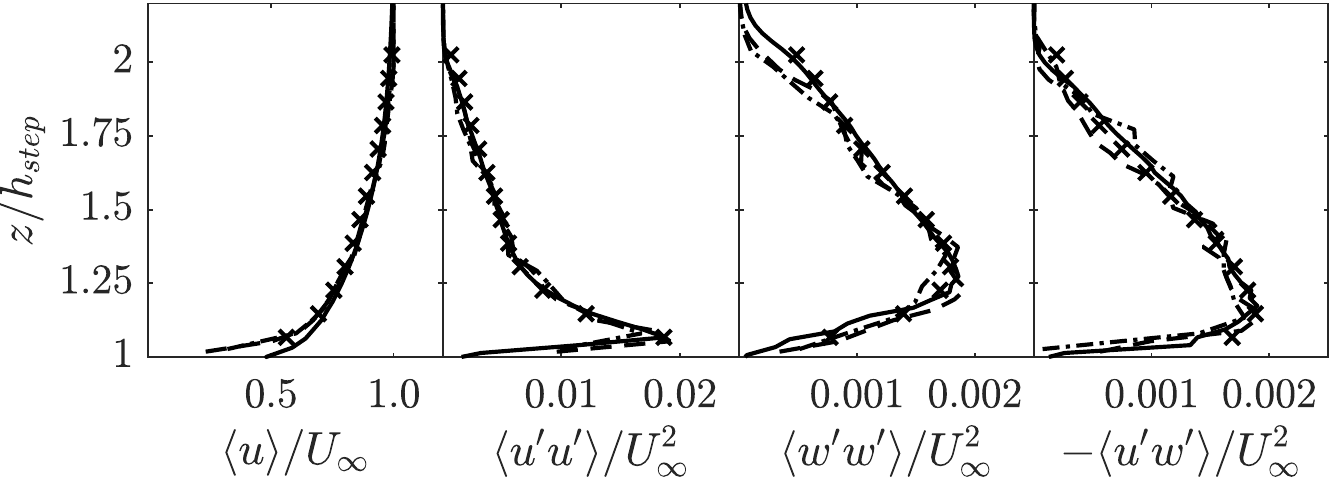}
 \end{center}
\caption{Turbulent statistics along the vertical direction from the precursor domain, using the proposed forcing method and three grid resolutions, sampled at the controlled forcing plane at conditions matching the momentum thickness Reynolds number of Jovic and Driver \cite{Jovic}. -.-.-) coarse; -\ -\ -) medium; -----) fine; X) experimental data}
\label{step_pre_grid}
\end{figure}

The mean streamwise velocity and non-zero Reynolds stress profiles sampled at the location of controlled forcing plane are plotted in figure \ref{step_pre_grid}. These profiles were collected from the precursor simulation using the proposed forcing method for three different grid resolutions and compared with the experimental data from Jovic and Driver \cite{Jovic}. As was seen in the high Reynolds number turbulent boundary layer case, the proposed forcing method was able to successfully match the target Reynolds stress profiles for all three grid resolutions.

\begin{figure}[h]
 \begin{center}
 Precursor) \\
  \includegraphics[width=12cm, clip=true]{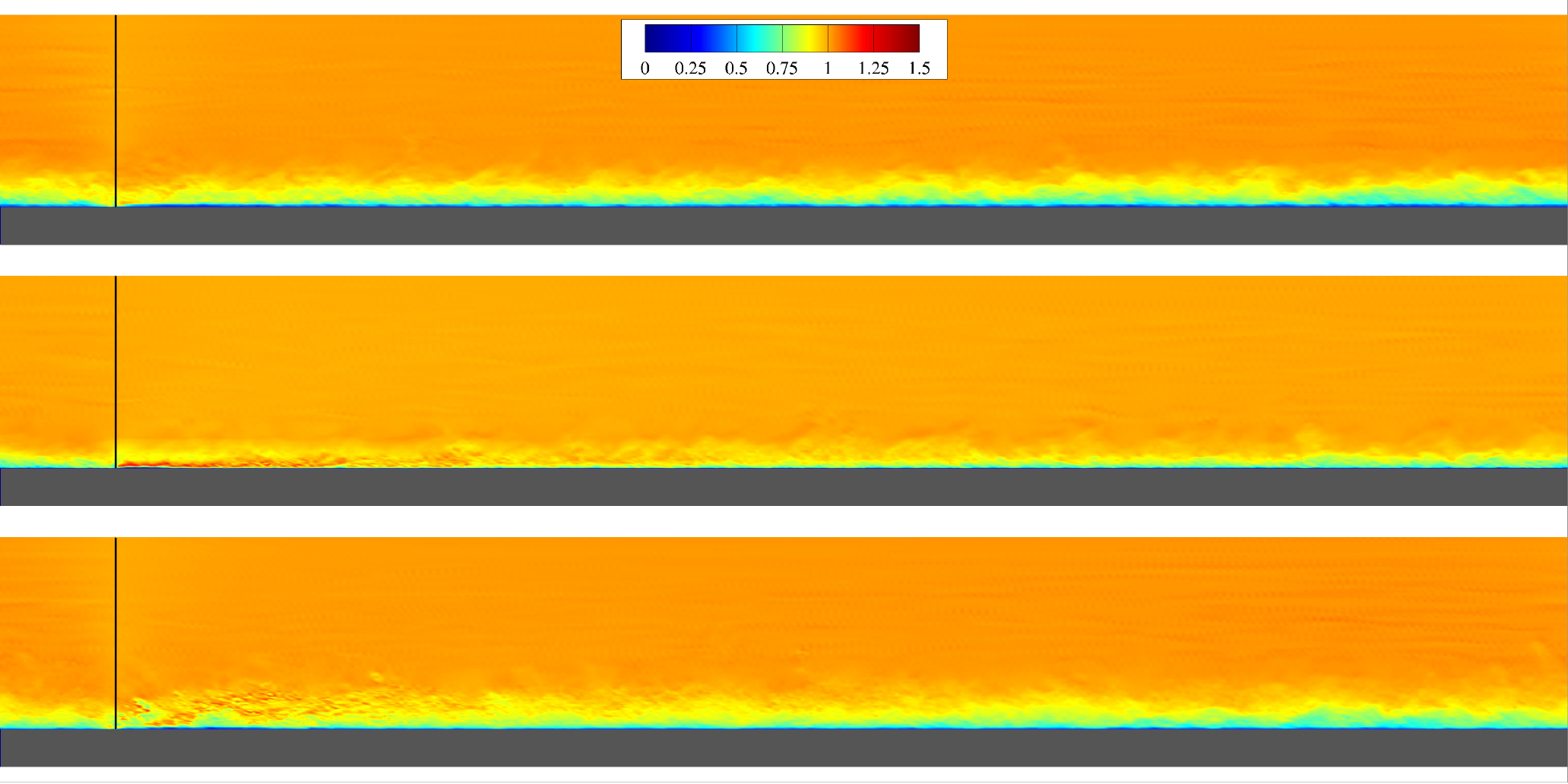} \\ \vspace{4mm} 
 Main) \\
  \includegraphics[width=12cm, clip=true]{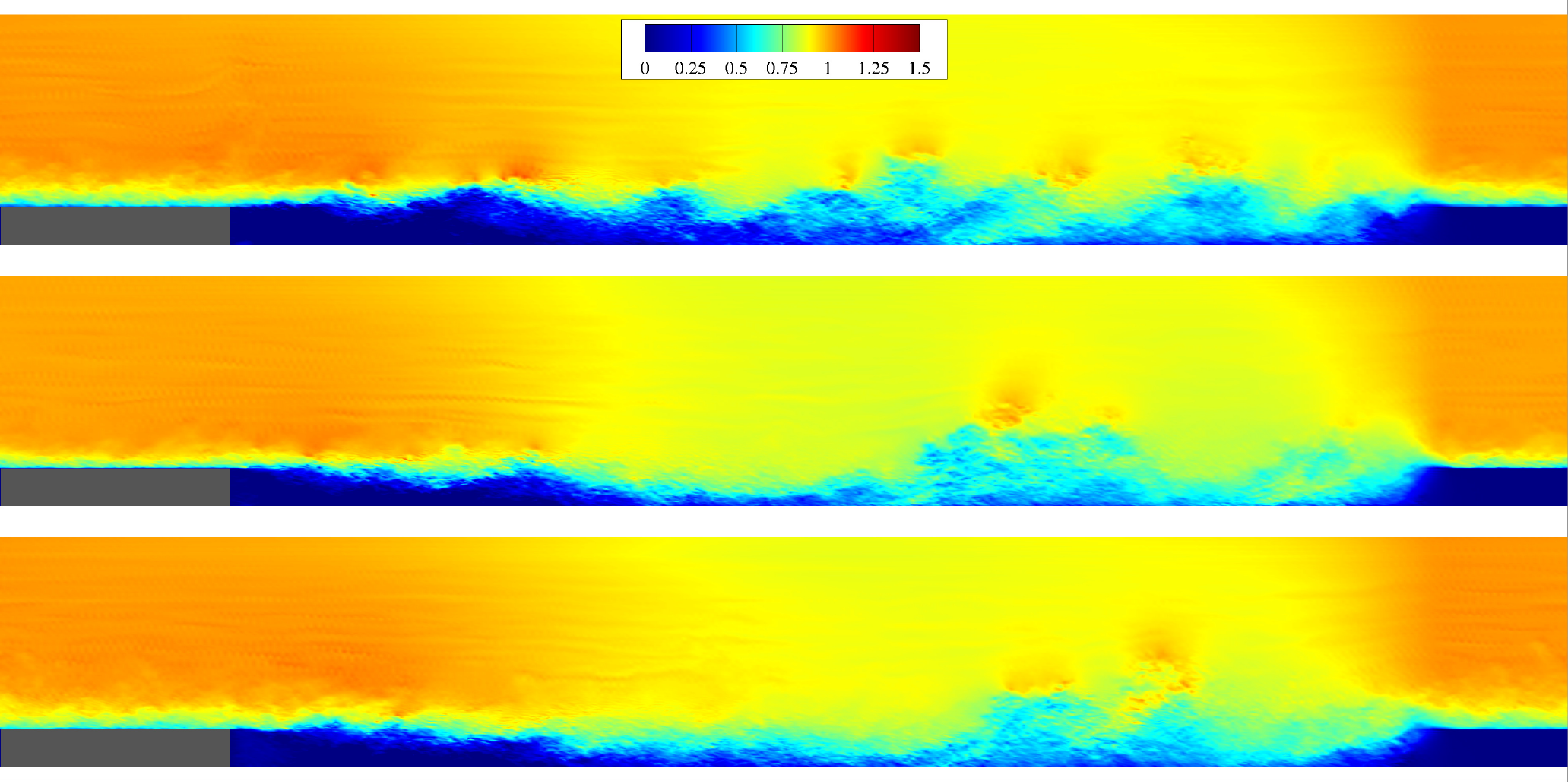}
 \end{center}
\caption{Contour of the instantaneous streamwise velocity normalized by the freestream velocity on an xz plane through the center of the domain, at conditions matching the momentum thickness Reynolds number in Jovic and Driver \cite{Jovic}. The black line represents the controlled forcing plane.} top 3) precursor domain; bottom 3) main domain; top) ``no forcing''; middle) ``original forcing''; bottom) proposed forcing
\label{step_u_contour}
\end{figure}

Contour plots of the instantaneous streamwise velocity in the precursor and main domain for the three different forcing cases are presented in figure \ref{step_u_contour}. The areas of grey represent where the Immersed Boundary Method is applied to model the presence of the step. The region of zero velocity immediately upstream from the outlet of the main domains is a product of the concurrent precursor forcing, clearly showing how the precursor forcing region successfully transfered the precursor flow field into the main domain. The highly turbulent wake of the step is smoothly transformed into the precursor domain flow. Looking at the precursor contour for the ``original forcing'' case in figure \ref{step_u_contour}, the region of much larger magnitude velocity near the wall immediately after the forcing plane is characteristic of the controlled forces still trying to converge. Whereas immediately downstream of the proposed forcing plane, there is only a small increase in velocity to counteract the natural decay as the fluctuations move throughout the rest of the domain.

\begin{figure}[h]
 \begin{center}
  \includegraphics[width=11cm, clip=true]{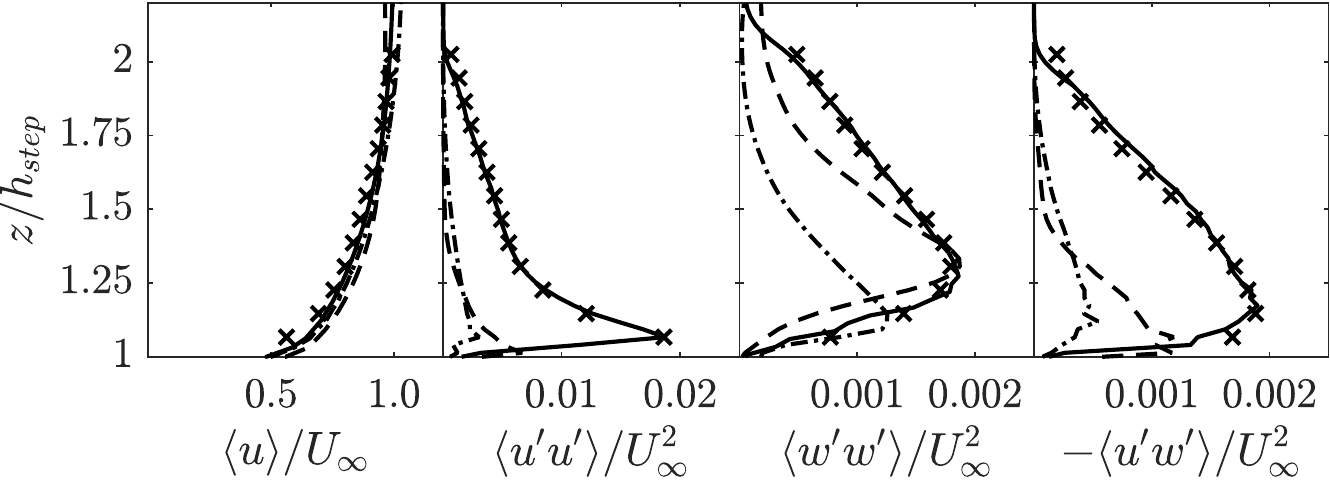}
 \end{center}
\caption{Turbulent statistics along the vertical direction from the precursor domain sampled at the controlled forcing plane at conditions matching the momentum thickness Reynolds number of Jovic and Driver \cite{Jovic}. -.-.-) ``no forcing''; -\ -\ -) ``original forcing''; -----) proposed forcing; X) experimental data }
\label{step_pre}
\end{figure}

The mean streamwise velocity and given Reynolds stress profiles from the precursor simulation are plotted in figure \ref{step_pre}, alongside the experimental data from Jovic and Driver at the location of the controlled forcing plane. Once again, excellent agreement was found for all of profiles for the proposed forcing case. Unlike the high Reynolds number boundary layer, the smaller boundary layer height allowed for the ``original forcing'' method to nearly converge on the wall-normal Reynolds stress by the time the proposed forcing reached convergence. This in turn produces a noticeable effect in the Reynolds shear stress, but only a small effect on the streamwise Reynolds stress at that point in time. The ``no forcing'' case also shows significant development in the wall-normal and Reynolds shear stress. The proposed forcing method is again seen to reduce the development time of the turbulent boundary layer as compared to the ``no forcing'' and ``original forcing'' cases.

\begin{figure}[h]
 \begin{center}
  \includegraphics[width=11cm, clip=true]{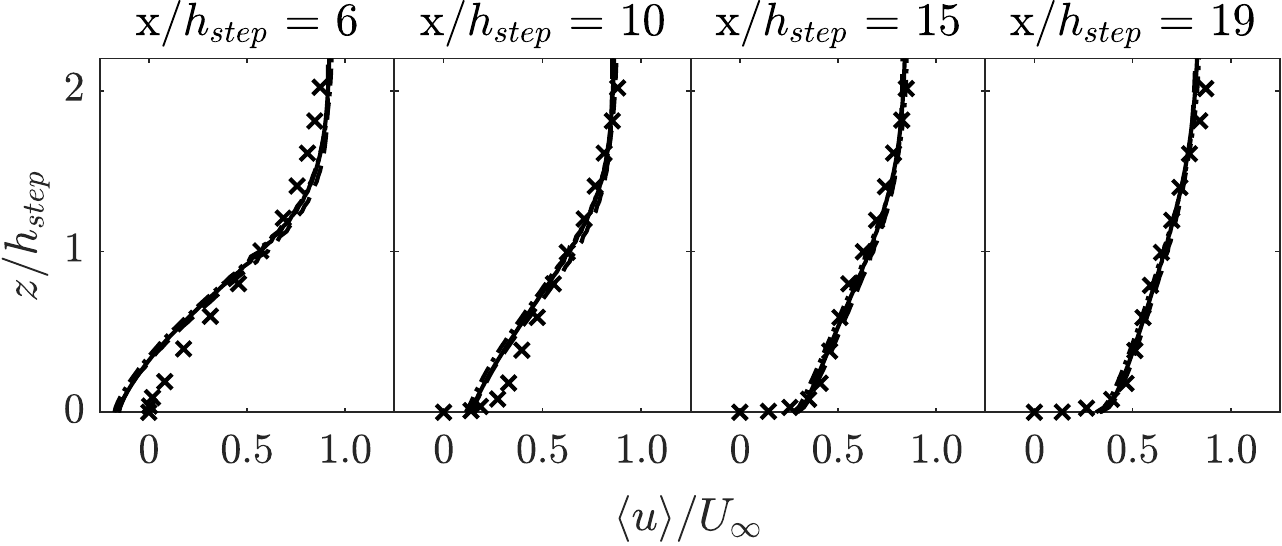} \\ \vspace{4mm}
  \includegraphics[width=11cm, clip=true]{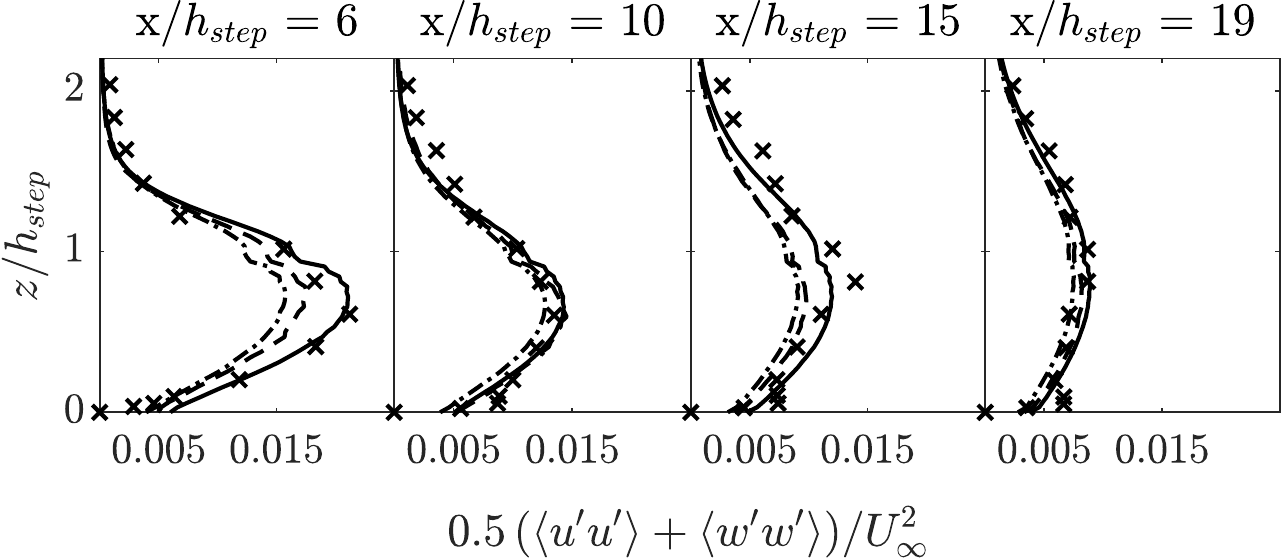} \\ \vspace{4mm}
  \includegraphics[width=11cm, clip=true]{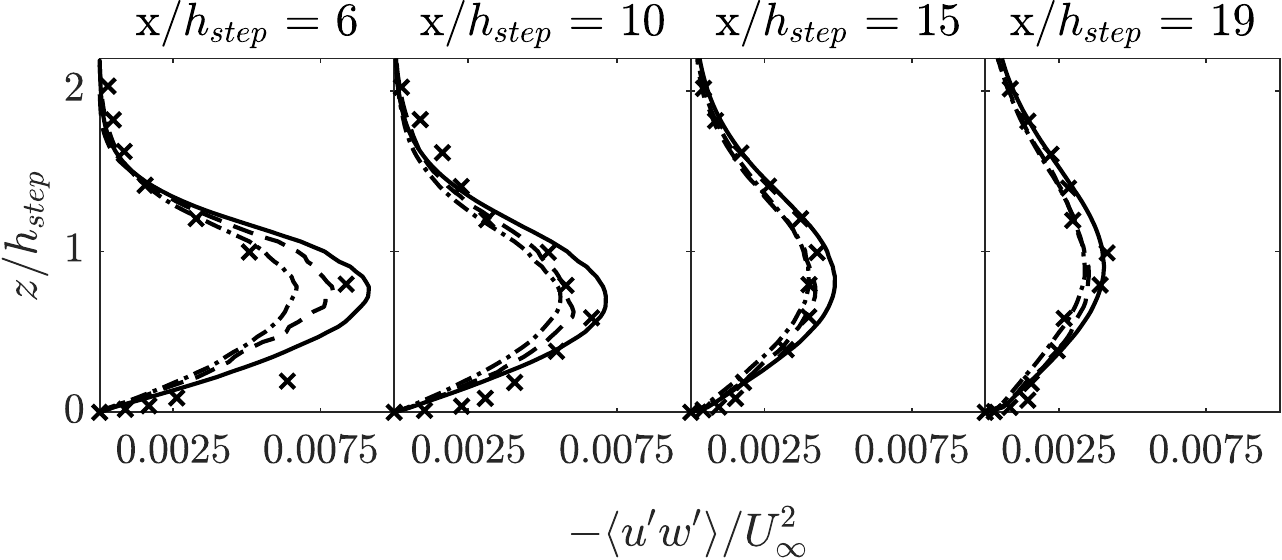}
 \end{center}
\caption{Turbulent statistics along the vertical direction sampled at streamwise locations downstream of the step. -.-.-) ``no forcing''; -\ -\ -) ``original forcing''; -----) proposed forcing; X) experimental data for the low Reynolds number case of Jovic and Driver \cite{Jovic}. top) Mean streamwise velocity; middle) 2D turbulent kinetic energy; bottom) Reynolds shear stress.    }
\label{step_main}
\end{figure}

The mean streamwise velocity, two-dimensional turbulent kinetic energy, and Reynolds shear stress profiles along the vertical direction are plotted in figure \ref{step_main} alongside the experimental data from Jovic and Driver at four streamwise locations downstream of the step, $x/h_{step}= \{6,10,15,19\}$. The profiles at $6h_{step}$ are located at the edge of recirculation bubble in the experimental flow. As seen in the mean velocity profile, the numerical simulations overpredicted the length of this bubble. This was expected because the Monin-Obukhov similarity theory used to model the flow at the wall was not developed for use in regions of recirculating flow. The wall model predicts the size of the recirculation bubble close enough to still allow comparisons of the turbulent statistics farther away from it. The farther away the profiles are from the recirculating region the better the solution calculated with the proposed forcing method matches with the experimental data. The ``no forcing'' and ``original forcing'' cases both show lower two-dimensional turbulent kinetic energy and Reynolds shear stress around one step height, with the difference between the proposed forcing case and the other two cases growing smaller moving downstream of the step. For this case, the imposition of the correct levels of boundary layer turbulence only gives a minor improvement.  

\section{Conclusion}
An extension of the original controlled forcing method was proposed and added into an existing concurrent precursor simulation method along with a mean flow forcing method. By calculating the controlled forces in the principal-axis coordinate system and applying them to each of the momentum equations, after a transformation, the Control Forced Concurrent Precursor Method is able to match a full anisotropic Reynolds stress tensor. Using high and low Reynolds number turbulent boundary layer flows as a test cases, the CFCPM matched the given mean velocity and both 3D and 2D Reynolds stress tensor profiles in the precursor simulation. The main domain flows showed good agreement with experimental wind-tunnel results for flows around a wall-mounted cube and over a backward-facing step. Simulations without forcing and with only the original controlled forcing did not reproduce the desired Reynolds stresses in the precursor simulations within the time period it took for the proposed controlled forcing to reproduce them. This shows that the proposed controlled forcing reduced the development times of the two turbulent boundary layers. The proposed controlled forcing also showed a modest improvement in agreement with the experimental results over the other two forcing cases in the main domain for the high Reynolds number turbulent boundary layer case, but only a slight improvement in low Reynolds number case.

\section*{Compliance with Ethical Standards}
\textbf{Conflict of Interests}\quad The authors declare that they have no conflict of interest.



\end{document}